\documentclass[aps,prb,twocolumn,superscriptaddress]{revtex4}
\usepackage[T1]{fontenc}
\usepackage[utf8]{inputenc}
\usepackage{newunicodechar}
\newunicodechar{ü}{\"{u}}
\newunicodechar{ö}{\"{o}}
\newunicodechar{Ö}{\"{O}}
\newunicodechar{ø}{\o}
\newunicodechar{á}{\'{a}}
\newunicodechar{Γ}{\Gamma}

\usepackage{graphicx}
\usepackage{hyperref}
\usepackage{graphicx}
\usepackage{color}
\usepackage[normalem]{ulem}

\begin{document}


\title{Synthesis of Epitaxial Single-Layer MoS$_2$ on Au(111)}


\author{Signe S. Gr\o nborg}
\author{S\o ren Ulstrup}
\author{Marco Bianchi}
\author{Maciej Dendzik}
\author{Charlotte E. Sanders}
\author{Jeppe V. Lauritsen}
\author{Philip Hofmann}
\email {philip@phys.au.dk}
\author{Jill A. Miwa}
\affiliation{Department of Physics and Astronomy, Interdisciplinary Nanoscience Center, Aarhus University, 8000 Aarhus C, Denmark}


\date{\today}

\begin{abstract}
We present a method for synthesizing large area epitaxial single-layer MoS$_2$ on the Au(111) surface in ultrahigh vacuum. Using scanning tunneling microscopy and low energy electron diffraction, the evolution of the growth is followed from nanoscale single-layer MoS$_2$ islands to a continuous MoS$_2$ layer. An exceptionally good control over the MoS$_2$ coverage is maintained using an approach based on cycles of Mo evaporation and sulfurization to first nucleate the MoS$_2$ nano-islands and then gradually increase their size. During this growth process the native herringbone reconstruction of Au(111) is lifted as shown by low energy electron diffraction measurements. Within these MoS$_2$ islands, we identify domains rotated by 60$^{\circ}$ that lead to atomically sharp line defects at domain boundaries. As the MoS$_2$ coverage approaches the limit of a complete single-layer, the formation of bilayer MoS$_2$ islands is initiated.  Angle-resolved photoemission spectroscopy measurements of both single and bilayer MoS$_2$ samples show a dramatic change in their band structure around the center of the Brillouin zone. Brief exposure to air after removing the MoS$_2$ layer from vacuum is not found to affect its quality.
\end{abstract}



\maketitle

\section{Introduction}
Many transition metal dichalcogenides (TMDCs) are comprised of 2D sheets held together by weak van der Waals interactions \cite{Geim:2013}. As for the analogous case of graphite and graphene, the properties of the TMDCs change in subtle but important ways when going from the bulk material to a single-layer (SL) \cite{Novoselov:2005b,Butler:2013, Rao:2014,Bradley:2015}. For example, MoS$_2$  which is one of the most studied TMDCs, has a direct band gap as a SL in contrast to the bulk, and correspondingly different optical properties \cite{Mak:2010, Splendiani:2010, Wang:2012, Rao:2013, Ramasubramaniam:2012, Ugeda:2014}. SL TMDCs can give rise to interesting new physics through the possibility of exploiting spin and valley degrees of freedom \cite{Xiao:2012, Xu:2014c,Zeng:2012,Zhang:2014c, Mak:2012}. They also exhibit promising chemical properties that are already employed in catalysis. MoS$_2$ and WS$_2$ nanoclusters, for example, are used for the desulfurization of fossil fuels \cite{Helveg:2000, Henrik2013}. 

Currently, the prevalent way to generate SL TMDCs is micro mechanical exfoliation and this has given rise to remarkable successes and structures that can be investigated primarily by transport and optical measurements \cite{Mak:2010, Splendiani:2010, Radisavljevic:2011, Xiao:2012, Cao:2012, Zeng:2012}.  An alternative to this approach is direct growth by chemical vapor deposition (CVD) \cite{Shi:2012} that can lead to large areas of high quality layers, as required for many experiments, and indeed for any large-scale fabrication of such materials. The synthesis of islands on the order of tens of nanometers grown by physical vapor deposition (PVD) has been reported for the (111) faces of Cu\cite{kimtoward2011} and Au.\cite{Sorensen:2014} However achieving  coverages in excess of 0.3\,ML without compromising the structural quality of the MoS$_2$ has been found to be difficult \cite{Sorensen:2014}. In this Article, we present a growth method based on reactive PVD that solves this problem and permits the growth of MoS$_2$ SLs with an almost unity coverage. The growth of bilayer (BL) MoS$_2$ islands is also achieved, which we use to compare the experimental electronic band structures of SL and BL MoS$_2$ on the Au(111) surface. A marked change in the valence band (VB) dispersion around the Brillouin zone (BZ) center at $\bar{\mathrm{\Gamma}}$ is found, pointing towards a more bulk-like character for the BL MoS$_2$ sample. We also show that these MoS$_2$ samples, grown under ultrahigh vacuum (UHV) conditions, are stable in air and atomically clean surfaces can be easily obtained by a mild anneal after re-introducing the samples into UHV. 

\section{Experimental}
\textbf{Detailed growth procedure of MoS$_2$ on Au(111).}
The synthesis of SL MoS$_2$ was performed in an UHV chamber (base pressure  of $<$2$\times$10$^{-10}$\,mbar) equipped with a home-built Aarhus-type scanning tunneling microscope (STM)\cite{Besenbacher}. With this experimental set-up, each step of the entire growth process could be monitored. All STM images were acquired with an etched W tip, and the bias voltages stated in the Article refer to the voltage applied to the sample.  The effect of instrumental artifacts, such as piezo creep, were minimized by calibrating the STM micrographs in the free WSxM software \cite{Horcas:2007} to match the known lattice constant of MoS$_2$ or the moir\'e periodicity of MoS$_2$ on Au(111), previously established in the literature \cite{Sorensen:2014}. The Au(111) sample was cleaned by repeated 2\,keV Ar$^{+}$ sputtering and 850\,K annealing cycles, and the cleanliness of the surface was verified by the presence of the Au herringbone reconstruction by STM. The synthesis was initiated by backfilling the chamber with H$_2$S gas to a pressure of 10$^{-7}$\,mbar using a custom-made doser with the end of the nozzle (4\,mm diameter) placed about 1\,mm away from the sample surface  to increase the local pressure of impinging H$_2$S gas \cite{Lauritsen:2006}.  Inside the stainless steel nozzle an approximately 3\,cm long capillary with a diameter of 40\,$\mu$m was mounted to constrict the flow of H$_2$S from a high pressure reservoir to a high vacuum system.   In this H$_2$S atmosphere and with the sample held at room temperature, metallic Mo was deposited onto the sample surface using a commercially available e-beam evaporator (EGCO4, Oxford Instruments). The deposition rate in our experiment was calibrated by a separate deposition of Mo onto a clean Au(111) surface and was found to be $\sim$0.08 ML/minute; this deposition rate was used throughout the experiment.  After a 5\,minute Mo deposition, the H$_2$S atmosphere was maintained while the sample was annealed to 850\,K for 30\,minutes. Once the sample reached a temperature of 450\,K during the subsequent cooling, the H$_2$S gas flow was stopped and the gas was pumped out of the chamber. These two steps -- i.e Mo deposition and anneal, which both take place in the H$_2$S environment -- constitute one \textit{growth cycle}. The first growth cycle yields a coverage of up to 0.3\,ML. We note that during the anneal step some of the deposited Mo may have alloyed with the Au crystal thereby reducing the expected MoS$_2$ coverage per growth cycle \cite{PRBChristiensen}. The growth cycle is repeated until an almost continuous coverage of SL MoS$_2$ is obtained, each cycle increasing the overall size of the MoS$_2$ islands. We were unable to achieve high quality SL MoS$_2$ beyond a coverage of 0.3~ML \textit{without} cycling the growth process.  A single long Mo deposition in H$_2$S only produced low quality sub-ML films. Furthermore, cooling the sample to 300\,K between each cycle appeared to promote the growth of high quality films. With this synthesis method, the MoS$_2$ coverage can be easily tuned to accommodate the experiment or application at hand.

We found that once a coverage of $\approx$0.8\,ML was reached, regions of BL MoS$_2$ began to seed and form.  Increasing the number of growth cycles produced more and larger islands of BL MoS$_2$, whose subsequent growth and evolution were similar to the growth and evolution of SL MoS$_2$; however, the contours of the BL MoS$_2$ islands were noticeably more irregular compared to the SL MoS$_2$ islands at a similar coverage. 

\textbf{Angle-resolved photoemission and low energy electron diffraction measurements.}
Angle-resolved photoemission spectroscopy (ARPES) measurements were performed at the SGM-3 beamline \cite{Hoffmann:2004} at the ASTRID2 synchrotron radiation source. A photon energy of 70\,eV was used to measure the electronic band structure of SL and BL MoS$_2$, as this energy showed the strongest enhancement of the layer-dependent changes in the electronic states \cite{Miwa:0}. The energy and momentum resolution were 25\,meV and 0.02~\AA$^{-1}$, respectively. Low energy electron diffraction (LEED) patterns were acquired for MoS$_2$-covered and clean Au(111) surfaces using a LEED optics mounted on the ARPES analysis chamber. The base pressure in the analysis chamber was in the low 10$^{-10}$\,mbar regime, and the sample temperature was held at 70\,K for the duration of the measurements.

\section{Results and Discussion}

\begin{figure*}
\includegraphics[width=16.4cm]{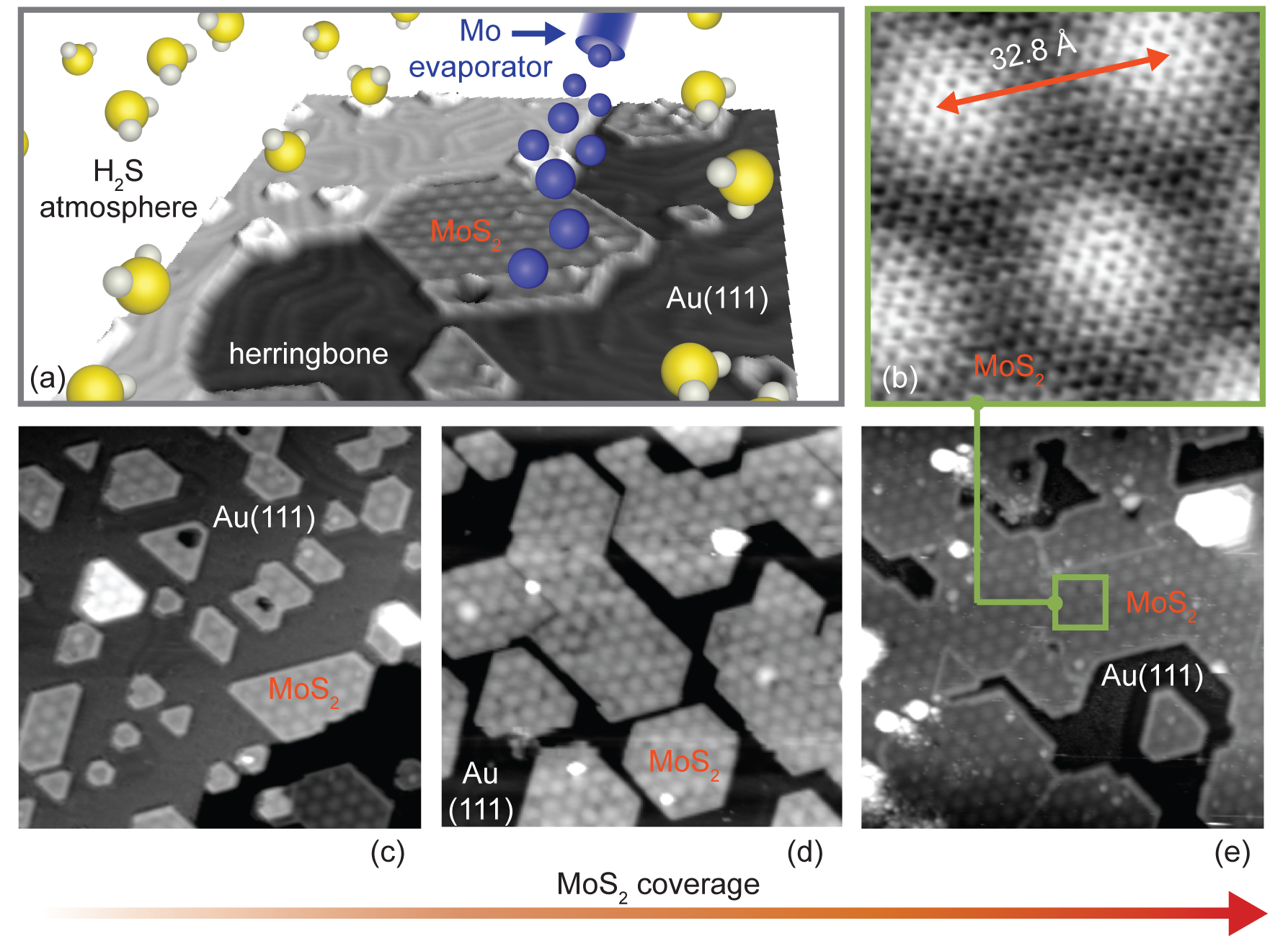}\\
\caption{(a) Schematic depicting the process by which SL MoS$_2$ nanoscale islands are grown on Au(111). Mo is evaporated onto the Au surface in a H$_2$S atmosphere. Following Mo deposition, the surface is annealed while the H$_2$S pressure is maintained.  (b) High resolution STM image of the moir\'e lattice of MoS$_2$ within the area demarcated by the green square in (e). The double headed arrow marks the moir\'e periodicity. (c) STM image acquired after one growth cycle showing hexagonal islands of SL MoS$_2$ (coverage is $\approx$0.3\,ML). Continued growth leads to coalescence of islands and a gradual increase in coverage as shown for (d) $\approx$0.6\,ML and (e) $\approx$0.8\,ML. The STM images are 65\,nm$\times$65\,nm except for the atomically resolved STM image in (b) that is 6\,nm$\times$6\,nm. Imaging parameters are approximately  1.13\,nA  and -0.15\,V. }
\label{fig:1}
\end{figure*}

The synthesis of large-area SL MoS$_2$ has been systematically explored by varying the Mo deposition rate, H$_2$S pressure and dose, sample temperature and annealing time. Overall we find that the successful synthesis of high-quality extended SL MoS$_2$ is strongly affected by two opposing effects concerned with the solubility of Mo in Au (alloy formation) and the diffusivity of Mo species on the surface. Both effects are strongly influenced by the sample temperature, but the deposition in H$_2$S serves to strongly suppress the alloy formation due to the reaction between the Mo and S, as well as the binding between S and Au. When the Mo deposition is carried out in a $\sim$10$^{-7}$\,mbar H$_2$S atmosphere and subsequently post-annealed at a relatively low temperature (673\,K), the result is a very well-defined array of triangular MoS$_2$ nanoparticles as reported in Ref. \cite{Helveg:2000} . This interesting MoS$_2$ morphology is attributed to a relatively low mobility of Mo at this temperature and the fact that nucleation of Mo predominantly occurs at the elbows of the characteristic Au(111) herringbone reconstruction \cite{Barth:1990}, yielding extended ordered arrays of amorphous Mo clusters on the surface. The higher post-anneal temperature of $\approx$850\,K used here for a duration of 30 minutes in the H$_2$S environment increases the mobility of Mo on the surface and promotes the growth of SL MoS$_2$ islands that are tens of nanometers in size\cite{Sorensen:2014}. Alloying of Mo and Au at these elevated temperatures would be very severe, but the effect is strongly reduced due to the presence of the H$_2$S atmosphere. Choosing this set of experimental parameters leads to relatively large MoS$_2$ islands with truncated triangular and hexagonal shapes on the Au(111) surface as observed in the STM image in Fig. \ref{fig:1}(c). Such islands are oriented with the (0001) basal plane of MoS$_2$ parallel to the Au surface and are seen to be distributed over the entire Au surface, which in this case reflects a total SL MoS$_2$ coverage of $\approx$0.3\,ML. The high resolution STM image of the area marked by the green square in Fig. \ref{fig:1}(e) reveals the atomically resolved lattice and the moir\'e structure of MoS$_2$. The incommensurability of the Au lattice (lattice constant of 2.88~\AA) and MoS$_2$ lattice (3.15~\AA) gives rise to a moir\'e structure that manifests itself on the (0001) basal planes of MoS$_2$ as clearly visible protrusions in a hexagonal pattern with a periodicity of 32.8~\AA~ (denoted by the double-headed orange arrow in Fig. \ref{fig:1}(b)).  Further attempts to increase the size of these islands by increasing the Mo coverage did not lead to a more uniform SL MoS$_2$ coverage and it was not possible to increase the annealing temperatures above 850\,K due to severe alloying of the Mo and Au that occurs irrespective of the H$_2$S pressure. Instead of increasing the temperature and Mo coverage in one step, the key to increasing the coverage of high quality SL MoS$_2$ involves repeated cycles of Mo evaporation and annealing in H$_2$S, using the aforementioned conditions. This results in islands merging together at a MoS$_2$ coverage of $\approx$0.6\,ML as shown in Fig. \ref{fig:1}(d). Continued cycling of the growth process leads to a further increase in coverage ($\approx$0.8\,ML)  and the formation of a nearly contiguous SL of MoS$_2$,  see Fig. \ref{fig:1}(e). 

\begin{figure} 
\includegraphics[width=8.2cm]{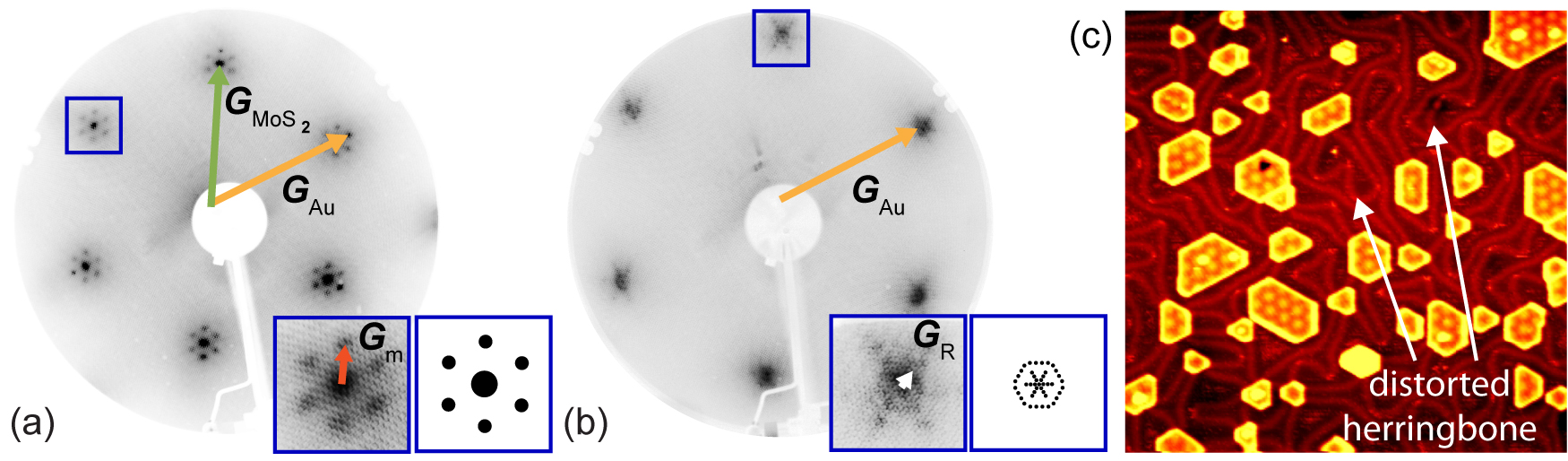}
\caption{LEED patterns acquired using electrons with a kinetic energy of 114\,eV for (a) a high coverage ($\approx$0.8\,ML) MoS$_2$ on Au sample and (b) a clean Au(111) surface.  The Au, MoS$_2$ and  moir\'e  reciprocal lattice vectors are shown. In panel (a) the MoS$_2$ on Au sample reveals six sharp first order spots in a hexagonal pattern and around each of these spots are six satellite spots also arranged in a hexagonal pattern. The inset shows the details of one first order spot along with an illustration to clarify the position and orientation of the satellite spots with respect to the first order spot.  In panel (b) a similar LEED image is shown for the clean Au(111) sample.  The inset shows one of the first order spots and a schematic is provided to highlight the multitude of diffraction spots that arise because of the herringbone reconstruction. (c) STM image of the surface after one growth cycle of MoS$_2$ in which distortions of the native herringbone are already visible.  Image parameters: 100\,$\times$100\,nm, 0.65\,nA, -1.25\,V.}
\label{fig:2}
\end{figure}

The structural details of the $\approx$0.8\,ML coverage MoS$_2$ sample are studied by STM in Fig. \ref{fig:1}(b) and (e) and by LEED in Fig. \ref{fig:2}(a). A sharp hexagonal diffraction pattern with satellites surrounding each main diffraction spot is observed, suggesting a clean, ordered and extended SL of MoS$_2$. The satellite spots are consistent with the moir\'e pattern observed in the STM images, indicating a large-area presence of this pattern. To highlight the origin of each diffraction spot, we have added reciprocal lattice vectors in Fig. \ref{fig:2}(a). \textbf{\emph{G}$_{\mathrm{MoS_2}}$} points to the main diffraction spots corresponding to the MoS$_2$ lattice while the satellites are associated with the moir\'e  reciprocal lattice vector,  \textbf{\emph{G}$_{\mathrm{m}}$}, as well as the Au(111) reciprocal lattice vector \textbf{\emph{G}$_{\mathrm{Au}}$}.  A separate LEED image of the clean Au(111) surface is shown for comparison in Fig. \ref{fig:2}(b). The structure and diffraction pattern of the Au(111) surface reconstruction is well-established \cite{Barth:1990,Huang:1990,Sandy:1991,Hove:1981}, but we summarize it here to facilitate a comparison with the diffraction pattern of the MoS$_2$ adlayer. The LEED image of Au(111), acquired at 300\,K, also exhibits a sharp hexagonal diffraction pattern with accompanying hexagonally positioned satellite features because of the rectangular 22$\times$$\sqrt{3}$ Au reconstructed unit cell which encompasses both face-centered cubic (fcc) and hexagonal close-packed (hcp) regions that arise from an average 4.55$\%$ compression of the first layer along the [1$\bar{1}$0] directions of the bulk truncated surface \cite{Barth:1990, Iski:2012}. Ridges form between the two different close-packed regions, creating a chevron or herringbone superstructure with 120$^{\circ}$ rotational symmetry.  The implications of this superstructure are a multitude of extra spots amongst the satellite spots \cite{Huang:1990, Sandy:1991}.  As seen in the insets of Fig. \ref{fig:2}(b), these extra spots define the outer contour of the small hexagon and radiate out along spokes from its center.  The hexagonal arrangement of the satellite spots for the clean Au(111), compared to the high coverage MoS$_2$ sample, are rotated with respect to one another by 30\,$^{\circ}$ -- see insets of Fig. \ref{fig:2}(a) and (b) -- indicating that the satellite features from the clean Au sample result from the herringbone surface reconstruction while the satellite features for the high coverage MoS$_2$ sample are instead due to the moir\'e (i.e. the interaction between the MoS$_2$ and the underlying Au).  The reciprocal lattice vector for the reconstructed Au(111) surface,  \textbf{\emph{G}$_{\mathrm{R}}$}, is shown in the inset of Fig. \ref{fig:2}(b) and the magnitude of the vector is scaled with respect to  \textbf{\emph{G}$_{\mathrm{m}}$} in the corresponding inset of Fig. \ref{fig:2}(a).  We note that higher order spots associated with the Au herringbone structure are also visible in the LEED pattern of Fig. \ref{fig:2}(b).  Since both diffraction patterns are acquired using electrons with a kinetic energy of 114\,eV, the absence of the Au satellite features in the  high coverage MoS$_2$  LEED data of Fig. \ref{fig:2}(a) strongly suggests that the herringbone reconstruction is lifted upon addition of the MoS$_2$ adlayer. These same Au spots are absent in LEED measurements of the MoS$_2$ sample acquired at different kinetic energies. Such a lifting of the reconstruction is not surprising as there are numerous reported instances of the addition of adatoms and adlayers leading to a redistribution of substrate surface atoms; for example, small species such as H, CO, H$_2$S and O$_2$ will perturb the peculiar quasi-hexagonal reconstruction  of  Pt(001) \cite{Ulstrup:2013, Nilsson:2012}, and self-assembled networks of oxygen- and sulphur-containing molecules can partially or fully lift the herringbone reconstruction of Au(111)  \cite{Iski:2012, Gatti:2014,Jewell:2010,Poirier:1996,Poirier:1997,Sorensen:2014}.  Indirect evidence of this lifting is seen for a small number of MoS$_2$ islands that appear brighter than the rest and still display a moir\'e pattern, such as the brighter island in Fig. \ref{fig:1}(c). This increase in apparent height is exactly consistent with the 2.3~\AA~height of an extra layer of Au atoms beneath the SL MoS$_2$ island \cite{Jewell:2010}. Such small extra Au islands are  expected to be generated by the lifting of the herringbone reconstruction, as this reconstruction has a higher density of atoms (4.5\,$\%$) in the first layer compared to the truncated bulk surface \cite{Poirier:1997}. The lifting of the native Au reconstruction is also supported by the STM image in Fig. \ref{fig:2}(c) which shows triangular-loop-like distortions of the regular herringbone reconstruction even after only one MoS$_2$ growth cycle.  These perturbations are similar to previously reported STM data of Au thin films grown on Ru(0001) which suggest that such triangular-like structures are amenable to relieving strain in stacked hexagonal layers \cite{Ling:2006}.

\begin{figure} 
\includegraphics[width=8.2cm]{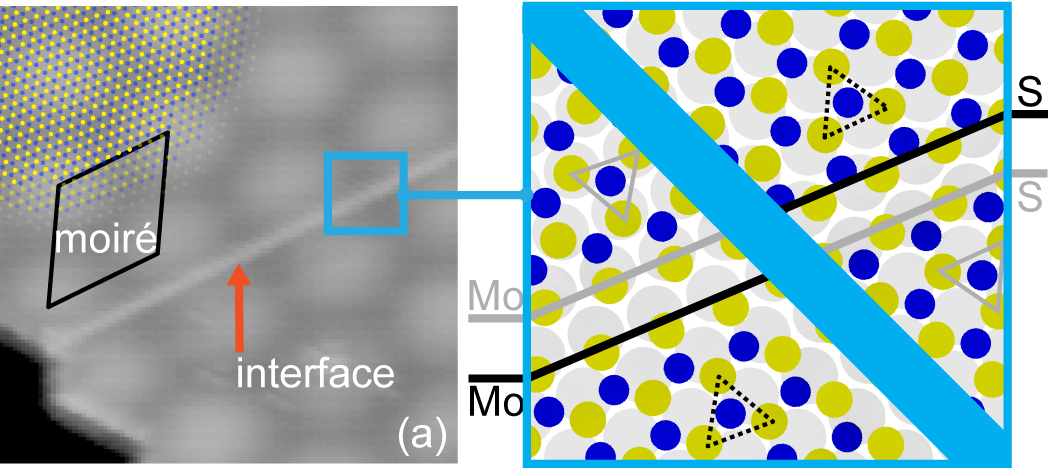}
\caption{(a) STM image of a line defect at the boundary between rotational domains. STM imaging parameters: 10\,nm$\times$10\,nm, 0.52\,nA and -0.88\,V. The moir\'e unit cell and ball model are overlaid to highlight the atomic character of the line defect. The area marked by the light blue square is represented by the ball model  at higher magnification in panel (b). (b) Schematic of the line defect arising from the coalescence of two oppositely oriented domains. The absolute orientation of the domains could not be determined from the STM data and therefore both possible scenarios for mating edges are shown and separated by the light blue diagonal line. The Mo-edge and S-edge are simply labelled Mo and S, respectively. The dashed (black) and solid (grey) outlined triangles are added to illustrate the two different rotational domains.  Ball model color code: yellow: S, blue: Mo, grey: Au.}
\label{fig:3ny}
\end{figure}

The STM images also reveal the presence of distinct line defects, as shown in Fig. \ref{fig:3ny}(a). These line defects are  boundaries between two opposite rotational domains of the three-fold symmetric MoS$_2$ lattice on the unreconstructed Au(111) surface lattice. Similar line defects have been observed in scanning transmission electron microscopy measurements of CVD-grown MoS$_2$ on SiO$_2$ \cite{Zhou:2013}.  The MoS$_2$ islands are either terminated in the $($10$\bar{1}$0$)$ or ($\bar{1}$010) directions. These terminations are referred to as the Mo-edge and S-edge, respectively.  Joining islands with the same domain orientation -- which is equivalent to the meeting of two different edge types -- will result in the formation of a continuous SL of MoS$_2$.  However, structural domain boundaries arise when domains of differently oriented MoS$_2$, or correspondingly the same edge types, meet during the growth process as shown in Fig. \ref{fig:3ny}. 

It is not clear from the STM images which type of interface is present for each individual line defect as only the upper layer of S atoms is imaged by STM \cite{kobayashielectronic1995}. Hence both possible interfaces are shown in Fig. \ref{fig:3ny}(b) separated by a light blue line. The  60$^{\circ}$ rotational domains are neither expected to influence the LEED pattern (although they would affect the relative spot intensities), nor would they have an effect on the overall electronic structure of SL MoS$_2$ due to the hexagonal symmetry of the reciprocal lattice.  However, the presence of line defects suggests that domains rotated by 60$^{\circ}$ coexist in our sample, and this is of crucial importance as it implies that the high symmetry points $\bar{\mathrm{K}}$ and $\bar{\mathrm{K}}^{\prime}$ in the BZ from different rotational domains coincide in laterally averaging techniques such as ARPES.

\begin{figure} 
\includegraphics[width=8.2cm]{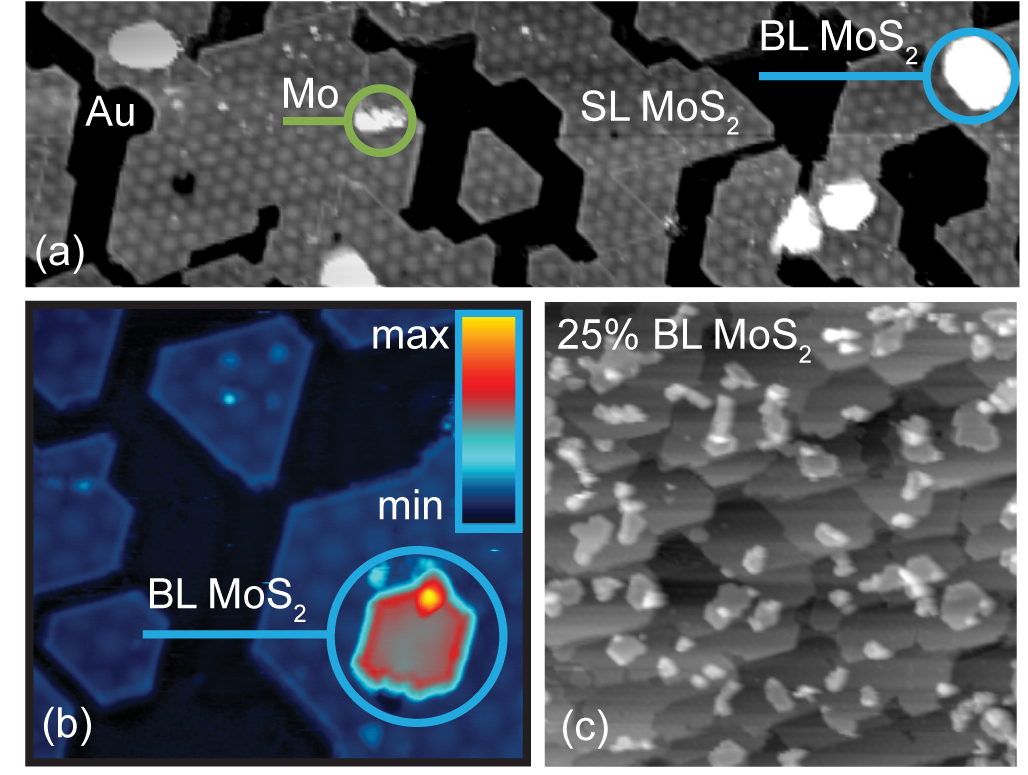}\\
\caption{(a) STM image of $\approx$0.8 ML MoS$_2$ on Au(111). Additional features are observed, specifically amorphous Mo clusters (green circle) and small islands of BL MoS$_2$ (blue circle).  Image parameters: 171\,nm$\times$49\,nm, 0.27\,nA and -1.23\,V.  (b) STM image highlighting the difference in apparent height of the BL MoS$_2$ compared to the surrounding  SL MoS$_2$ areas. Cool to warm colors represent low to high heights. Image parameters: 36\,nm$\times$34\,nm, 0.40\,nA and -1.16\,V. (c) STM image of MoS$_2$ on Au(111) with an estimated BL MoS$_2$ coverage of 0.25\,ML.  Image parameters: 175\,nm$\times$175\,nm, 0.19\,nA and -1.16\,V.  }
\label{fig:4}
\end{figure}

Our STM images of the high coverage SL MoS$_2$ films frequently reveal additional features atop SL MoS$_2$, which are evident from Fig. \ref{fig:4}. We assign the irregularly shaped clusters highlighted by the green circle in Fig. \ref{fig:4}(a) to amorphous Mo clusters, which may result from insufficient sulfurization during the growth cycles. Perhaps more interesting is the feature marked by the blue circle, which shows the formation of a BL MoS$_2$ island. A STM image of such an island is shown in Fig. \ref{fig:4}(b). In contrast to the SL islands, the basal plane of the BL appears flat in the center of the island; a moir\'e is absent and it is difficult to obtain atomic resolution. The apparent height of the BL islands relative to the underlying SL islands is $(5.2 \pm 0.3)$~\AA. This value is comparable to the bulk inter-layer spacing of 6.15~\AA~in the 2H-stacking of S-Mo-S layers, supporting the assignment of these features to BL and not multilayer islands. 

As the number of growth cycles is increased, we foster the formation of more and larger islands of BL MoS$_2$.  In the STM image shown in Fig. \ref{fig:4}(c), the BL regions appear brighter than the surrounding darker SL MoS$_2$ film.  Upon close inspection of the STM image, the SL MoS$_2$ still manifests a moir\'e pattern that is unaffected by the presence of the BL MoS$_2$ islands. Here, the BL MoS$_2$ covers an estimated $\approx$25\,$\%$ of the surface.   The contours of the BL MoS$_2$ islands are notably more irregularly shaped than SL MoS$_2$ islands for a similar coverage; see Fig. \ref{fig:1}(c) for comparison.  Such irregular contours have been observed for SL MoS$_2$ islands grown on highly ordered pyrolytic graphite \cite{Kibsgaard:2006} and more recently on graphene \cite{Miwa:0}, where the adlayer-substrate interaction is governed by van der Waals forces.  Given that, in BL regions, the interaction between the second layer of the BL MoS$_2$ and the underlying SL MoS$_2$ film is weaker than the interaction between the SL MoS$_2$ and the underlying Au substrate, we might expect to see growth of islands with many different rotational domains orientations.  This tendency to form various rotational domains of BL MoS$_2$, together with the irregular shapes of the BL islands, suggests that such islands may not seamlessly  merge together during  subsequent growth cycles and will indeed make the synthesis of large domains of high quality multilayers of MoS$_2$ challenging.

\begin{figure} 
\includegraphics[width=8.2cm]{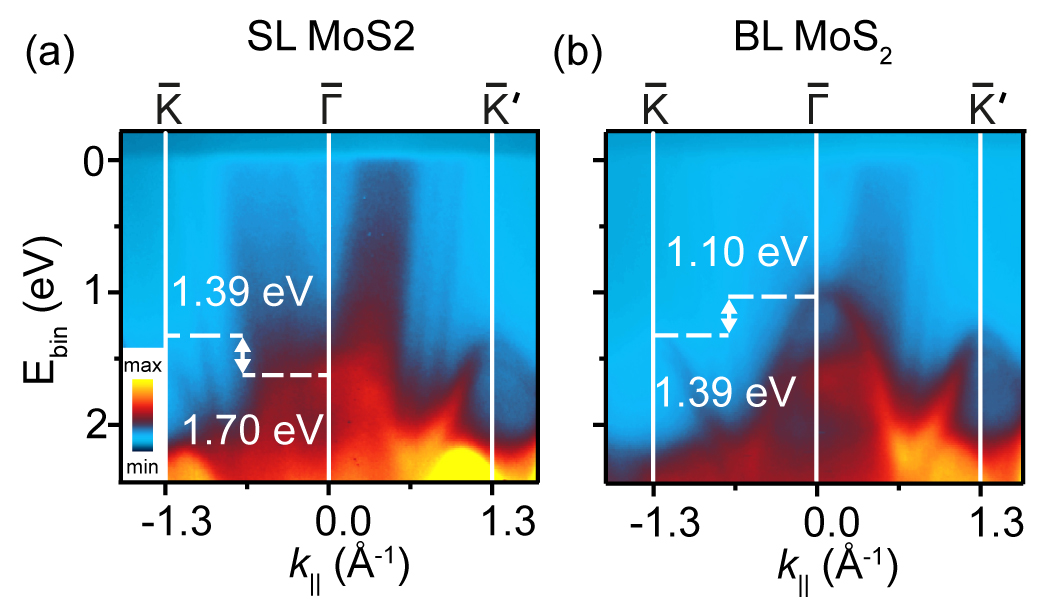}\\
\caption{Band structure measurements of (a) SL and (b) BL MoS$_2$ on Au(111) by ARPES with a photon energy of 70\,eV. The VB dispersion along the $\bar{\mathrm{K}}$ -- $\bar{\mathrm{\Gamma}}$ -- $\bar{\mathrm{K}^{\prime}}$ direction of the MoS$_2$ BZ is shown for both samples. The position of the VB maximum at $\bar{\mathrm{\Gamma}}$ compared to $\bar{\mathrm{K}}$ is highlighted in each case by the dashed white lines. }
\label{fig:5}
\end{figure}

We carried out electronic structure measurements by ARPES of both the SL and BL MoS$_2$ samples in order to compare their band structures on Au(111). The detailed features of SL MoS$_2$ on Au(111) in ARPES are discussed in Ref. \cite{Miwa:2014}. Here we focus on the VB dispersion along the $\bar{\mathrm{K}}$ -- $\bar{\mathrm{\Gamma}}$ -- $\bar{\mathrm{K}^{\prime}}$ direction of the MoS$_2$ BZ as shown in Fig. \ref{fig:5}. For SL MoS$_2$ the global VB maximum is situated at the $\bar{\mathrm{K}}$-point, which resides within the projected band gap of the Au electronic structure. Consequently the electronic states of MoS$_2$ in this region of the BZ do \textit{not} interact with the Au bulk states, and the MoS$_2$ bands appear sharp and well-defined \cite{Takeuchi:1991}.  The upper VB at $\bar{\mathrm{K}}$ exhibits a  spin splitting with a measurable separation of (145$\pm$4)\,meV \cite{Miwa:2014}.  The top of the VB at $\bar{\mathrm{\Gamma}}$ falls within the continuum of projected Au bulk states. The broad character of these VB states is attributed to their interaction with the Au bulk states \cite{Zhu:2011,Cappelluti:2013,Miwa:2014}. This observation is in contrast to ARPES data acquired for SL MoS$_2$ grown on graphene where the MoS$_2$ band structure is completely unperturbed by the substrate \cite{Miwa:0}.  

The most pronounced difference between the ARPES measurements of the SL and BL MoS$_2$ samples is the binding energy position of the VB maximum at $\bar{\mathrm{\Gamma}}$.  The binding energy differences with respect to $\bar{\mathrm{K}}$ are highlighted by the dashed white lines along with the salient binding energy, $E_{bin}$, values.  For SL MoS$_2$, the VB at $\bar{\mathrm{\Gamma}}$ exhibits a local maximum at a binding energy of 1.70\,eV, which is 0.31\,eV \textit{below} the VB global maximum (1.39\,eV) at $\bar{\mathrm{K}}$. This picture is consistent with the view that SL MoS$_2$ is a direct band gap semiconductor.  With ARPES only the occupied states of a material are probed, so this expectation cannot be immediately and directly verified.  However it has been previously shown that by alkali doping SL MoS$_2$ on Au(111), the conduction band minimum can be sufficiently occupied such that this direct band gap at $\bar{\mathrm{K}}$ can be measured \cite{Miwa:2014}.  For BL MoS$_2$ an extra dispersing band appears around $\bar{\mathrm{\Gamma}}$ at 1.10\,eV or 0.29\,eV \textit{above} the position of the VB local maximum at $\bar{\mathrm{K}}$.  (The positions of the VB maxima are determined from fits of the energy distribution curves which are not shown here but the method has been previously demonstrated for SL  MoS$_2$ \cite{Miwa:2014}.)  For BL MoS$_2$ the appearance of two distinct bands at $\bar{\mathrm{\Gamma}}$ (Fig. \ref{fig:5}a) arises from an interaction between the two MoS$_2$ layers, leading to the formation of a bonding/anti-bonding splitting of the VB \cite{Cheiwchanchamnangij:2012}.   This is a very distinct spectral feature of the BL case, and  in contrast to SL MoS$_2$ (Fig. \ref{fig:5}a) where the splitting of the bands at the VB maximum situated at $\bar{\mathrm{K}}$ is purely a spin-orbit effect due to a lack of inversion symmetry in the TMDC \cite{Zhu:2011}.  In fact, the direct to indirect band gap transition arising from additional interlayer interactions has recently been observed in optical spectroscopy measurements of MoS$_2$ samples with thicknesses ranging from one to six layers \cite{Mak:2010} and MoSe$_2$ with thicknesses up to eight layers \cite{Zhang:2014}. Furthermore, the MoS$_2$ bands around $\bar{\mathrm{\Gamma}}$ for BL MoS$_2$ appear a lot sharper and narrower than the bands near  $\bar{\mathrm{\Gamma}}$  for the SL case, something we ascribe to the increased interaction between the two layers rather than the hybridization of just a single adlayer with the underlying Au substrate. This is consistent with the appearance of the BL MoS$_2$ islands in STM, where they appear to be very weakly interacting with the substrate. 

\begin{figure} 
\includegraphics[width=8.2cm]{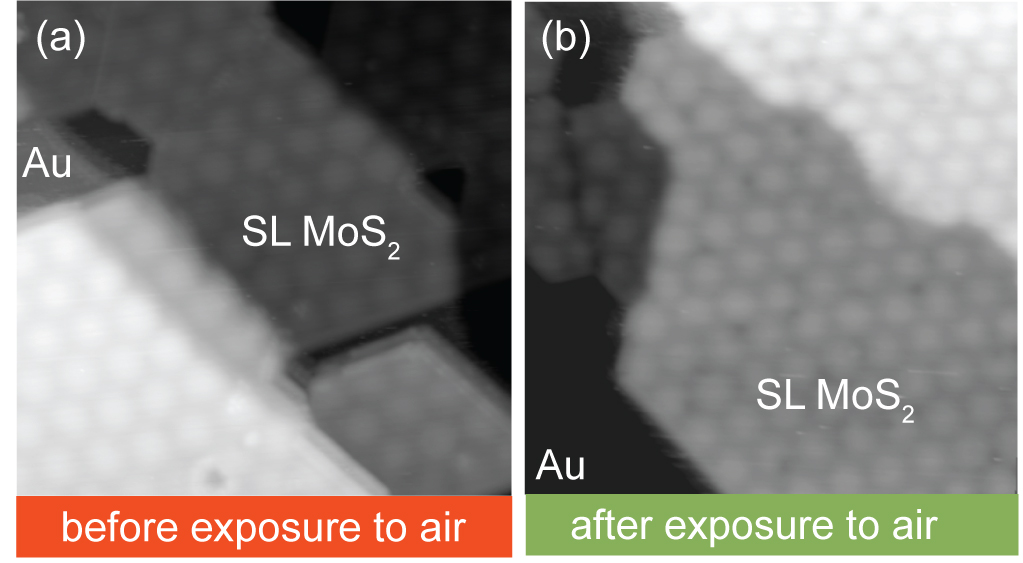}
\caption{STM images of SL MoS$_2$ samples acquired at room temperature (a) before and (b) after exposure to air.  The surface quality is essentially the same after the exposed surface is gently annealed to 500\,K. Imaging parameters: (a) 35\,nm$\times$35\,nm, 0.33\,nA and 0.78\,V.  and (b) 35\,nm$\times$35\,nm, 0.28\,nA and 1.00\,V.  }
\label{fig:6}
\end{figure}

The chemical stability of our epitaxial SL MoS$_2$ samples towards air exposure is tested as shown in Fig. \ref{fig:6}. The STM images present the SL MoS$_2$ lattice before (a) and after (b) air exposure at room temperature for a few hours, followed by a mild annealing to 500\,K in UHV to remove physisorbed species. Incidentally, these SLs appear to be stable up to even higher anneal temperatures of approximately 900\,K. Prolonged exposure to air, i.e. in excess of several hours, can lead to extensive accumulation of adsorbates, and these higher annealing temperatures are generally found to be more effective for cleaning the sample. The quality of the surfaces in the images is essentially identical, demonstrating  the chemical robustness of the epitaxial SL MoS$_2$ on Au(111). This result suggests an inert character of these SL MoS$_2$ samples towards an oxidizing gas -- a property which greatly simplifies the transfer of samples between different experimental set-ups.

\section{Conclusions}
We have developed a procedure for synthesizing large area epitaxial SL MoS$_2$ on Au(111), and demonstrated that these samples are robust against exposure to air.  We expect this UHV compatible growth method to further surface sensitive characterization studies and nanoscale device applications that may require larger areas of SL MoS$_2$ than those typically achieved by mechanical exfoliation. Moreover, a comparison of the ARPES data obtained from MoS$_2$ layers grown with this method \cite{Miwa:2014} to those from exfoliated samples \cite{Jin:2013aa}  suggest that the former may be of superior quality. Our MoS$_2$ synthesis method is based on cycles of Mo evaporation and sulfurization to gradually increase the coverage of SL MoS$_2$ and avoid excessive formation of amorphous Mo clusters and BL MoS$_2$ islands. The MoS$_2$ layer seems to be comprised of domains rotated by 60$^{\circ}$ which lead to the formation of straight line defects at domain boundaries. Upon comparison with LEED data of a clean Au(111) sample we find that the native herringbone reconstruction of Au is lifted with the addition of the MoS$_2$ adlayer. This controlled growth method of epitaxial SL MoS$_2$ should be transferable to similar TMDCs and other substrates as long as the interaction between the TMDCs and the surface is strong enough to seed initial island growth and prohibit the formation of random domain orientations. Rotational domains could be avoided by choosing surface orientations of reduced symmetry.

\section{Acknowledgement}
We gratefully acknowledge financial support from the VILLUM foundation, the Danish Council for Independent Research, Natural Sciences under the Sapere Aude program (Grant Nos. DFF-4090-00125 and DFF-4002-00029), the Lundbeck Foundation, the Danish Strategic Research Council (CAT-C) and Haldor Tops\o e A/S.

%
\end{document}